\documentclass[journal]{IEEEtran}

\usepackage{graphicx}
\usepackage{multirow}
\usepackage{url}

\begin{document}

\title{Collaborative peer production as an alternative to hierarchical internet based business systems}

\author{Ganesh.G $^{a,b}$, Debanjum Singh Solanky $^{a,c}$ \& Govindaraj.R $^{a,d}$ 
	
	\thanks{$^{a}$Central Electronics Engineering Research Institute (CEERI),
		Chennai Centre, CSIR Madras Complex, Taramani, Chennai-600113, Tamil Nadu, India.
	}
	
	\thanks{$^{b}$Research Intern, e-mail: 31gane@gmail.com.}
	
	\thanks{$^{c}$Student Trainee, email: debanjum@gmail.com}
	
	\thanks{$^{d}$Corresponding author, e-mail: rajagovind@hotmail.com}%
	
}


\maketitle

\begin{abstract}
	As we move towards more data intensive, device centric global communication networks, our ability to usefully harvest these large datastores is degrading. The widening asymmetry in the explosive growth of data versus our ability to use it, is forcing us towards centralized analytics. This splintered concentration of data further consolidates analytical capabilities in the hands of the few and divides the network into the analysors and the analysed. The fracturing of the system into opaque datastores and analytics blocks creates a strong positive feedback loop and has a significant negative impact on the stability, transparency and freedom of the network. This paper attempted to identify problems associated with the internet, internet dependent business models and reviewing available solutions and discuss possible solutions which became necessary.
\end{abstract}

\IEEEpeerreviewmaketitle

\section{Introduction}
The Internet is and has always been a highly effective medium and brewing platform for sharing - data, information and analytical tools along with the knowledge gained while creating them. With its rapid spread and adoption by the general populace its effectiveness has only increased and revolutionized knowledge sharing in the way which was not possible before. Evolution in Information and Communication Technology(ICT) with internet has provided ways to start business by utilizing its simplest and powerful tools which was not possible with earlier technological revolutions. This phenomenon has given rise to a number of different organizational strategies to use internet as business platform and resulted in the construction of several web based business models that has contributed to the commercialization in the internet\cite{rappa2004utility}.

It is a common practice to blend different models by commercial organizations in its overall Internet business strategy as explained in\cite{rappa2004utility}. The impetus for blending such different strategies together depends upon the products produced, market share of the product, financial throughput from sales and marketing. By inducing blend of strategies the organization can lay multiple monetizing routes in order to increase their profit margin. This has paved the way to utilize, contribute, adapt the web technologies for e-commerce to accelerate centralized business strategies. 
Such practice have catalyzed the transformation of internet from its initial democratic, collaborative information and resource sharing structure, into a monetizing platform for manufacturing, packaging, and marketing industry.

Table I refers to the state of the internet based business models as listed by Michael Rappa\cite{rappa2004utility}, along with the structures with which each model have developed and become dependent upon. The structure columns show how the current structure of internet was qualitatively classified as viewed by the internet based business models. As the table elucidates, most of the model have enriched and evolved with a centralized structure and influenced by the policies associated with it.

\begin{table}[h]
	\renewcommand{\arraystretch}{1.2}
	\caption{Internet based Business Models and their Structure}
	\begin{center}
		\begin{tabular}[m]{ |l|c|c|c| }
			\hline \multirow{2}{*}{Business Model}    & \multicolumn{3} { |c| } {Structure}  \\
			\cline{2-4}                               &   Centralized     &   Distributed     &   Collaborative     \\    
			\hline                  Brokerage         &   --              &   --              &   --                \\    
			\hline                  Advertising       &   Yes             &   --              &   --                \\    
			\hline                  Infomediary       &   Yes             &   --              &   --                \\
			\hline                  Merchant          &   Possible        &   --              &   --                \\
			\hline                  Manufacturer      &   Yes             &   --              &   --                \\
			\hline                  Affiliate         &   --              &   --              &   --                \\
			\hline                  Community         &   Possible        &   Yes             &   Yes               \\
			\hline                  Subscription      &   Yes             &   --              &   --                \\
			\hline                  Utility           &   Possible        &   Possible        &   Possible          \\
			\hline
		\end{tabular}
	\end{center}
\end{table}

Web technology is capable of facilitating whoever using internet with their own channel of communication which was not made possible with previous technologies. It has given an individual freedom to express their thoughts, share the created content and other digital resources through their personal computer and the internet. Business models which leveraged on the same technology for esatablishing their market share have used the information in the internet sprouting different information based businesses. Inherent competition between the organizations, made them dependent upon the information available in the internet and exploiting the same to reveal the patterns by investing more and more computational resources. This phenomenon has transformed many internet based business into information based business. So the devised business models rely heavily upon the information and processing the same to occupy more and more market share that drives the organizations to acquire as much information as possible from the consumers, searching for the pattern in the data that may help their business to sustain amidst the competition. 

Demonstrating a substantial increase in their profit margins this phenomenon has become ubiquitious practice among the organizations which influenced the new crowdsourced startup organizations like tellspec, scio, healbe to follow the same or similar model. This clearly shows that the business models and associated policies which biased more towards the information to increase the profit margin have recursively influenced the whole business style.

Symbiotic sharing of data, hardware and software design, between the peers in homebrewing, research and development communities are very vital in mutual development and contribution to the technology innovation. Communities approach follows the DIY\textit{(do it yourself)} and DNRY\textit{(do not repeat yourself)} methods in order to solve the problems themselves for which a prior solution is not available and to reduce the repetitive work by reusing the already exisiting peer reviewed solutions resulting in focused research for the unsolved problems. This recursive development strategy followed by most of the communities depend upon the distributed sharing of resources(information, designs, tools, computation). Business driven centralization of information accessibility and availability may produce adverse side effects on homebrewing, research and development communities in most of the fields that continuously develop and share -- tools and information developed by other peer members\cite{mackinnon2012consent}. Moreover, with the inclusion of web based remote services that highly depend upon the commons data, are offered back for a fixed cost\cite{handleyp2p,pasquinelli2009google,jakobsson2010pirates}. This has transformed commons and research communities into consumers of service market. 

One can understand how business policies and uncertainities associated with it can defeat the very purpose of the service, even when there is nothing wrong in its implemented mechanism\cite{levy2011plex}. It demonstrates how mechanism with which the web works can be controlled by the policies built to catalyze the profitization of the business. 

This paper provides a overall view about how the internet was used as a commercialization platform that fueled the transformation of internet to a platform populated with centralized structures. It also elucidates how monetization policy can shear the underlying mechanism deviating from achieving its original goal. The objective of the paper is to propose a conceptual solution implemented to a certain level which addresses the problems associated with peculiarities and uncertainities of the internet based business models.

The rest of the paper is structured as follows. Section II lists out the problems associated with how business models have contributed in recasting the internet, enriching it as a primary platform for e-business to grow while indirectly deprecating its free medium form. Section III describes the collaborative peer production model which was practiced and evolving with its own technological innovations and uncertainties. It also emphasizes about the need to generate new business models, that balances the economic incentives and free nature of internet. Concluding remarks is given in section IV.

\section{Characteristic problems associated with Internet based Business Models}

Eventhough internet based business models have intensified the competition for market share between the leading organizations offering their services, it has deprecated the internets ability as free communication medium between each individual users and communities.

To understand the characteristic problems, we shall assume a personal computer connected to internet at the user end as a node that constitutes enough computing power for performing three main operations - Data handling(acquisition and analysis), Human Machine Interaction, Communication and Digital Resource Sharing respectively. Considering such a node will help us better understand the real nature of the problem and its impact in possible directions. 

This section lists the characteristics and problems associated with the business models, directed in the increasing order of their impact on other fields and frameworks which depend upon internet for its functionality.

\subsection{Hidden perspectives}
Several perspectives would be necessary, to understand a business model that thrives on the inter-networking facility. Many reports have been published analyzing such business models related to specific market along with how new technological innovations can influence their business strategy with its uncertainities and peculiarities.

The importance of analysing different perspectives and their interrelations that are necessary to understand the mobile business market is demonstrated\cite{camponovo2003business}. Even though such analysis can help exploit the market from business point of view, nobody really knows how much the model and its pragmatic implementation is stressing out the complete framework upon which the model itself is built and working on. 

Internet is completely made by the individual users, communities and service providers which we refer here as players of internet. The balanced scorecard depicted in Fig.1 shows the perspectives taken into account for analysis by Giovanni and represents the user freedom and collaboration perspectives which are needed to be included for analysis. If the perspectives are analyzed without including the interrelations between such players, then the results of analysis may not contain parameters explaining the stress pressed upon the framework. Thus analysis requires two more perspectives which explains the influence of the players such as the users perspective and the social perspective.

Users perspective concentrates on their ability to control what they have purchased, from the market and not the other way which usually happens with centralized and proprietary business. This is the core philosophy with which the free software products are being developed and proved to have good quality, diversity and market share competing the proprietary products sold in the internet. This became possible because those products respect users freedom and allows them modify it as per their requirements and wishes. Such model have already been much recognized in educational, research communities and continues to expand in every possible domains. It is known for demonstrating alternatives for proprietary softwares that obscures mechanism and knowledge that works behind the tool. This gives rise to social perspective that concentrates on collaborative involvement and development of necessary tools required for the community.

\begin{figure}[h]
	\centering
	\includegraphics[width=0.90\linewidth]{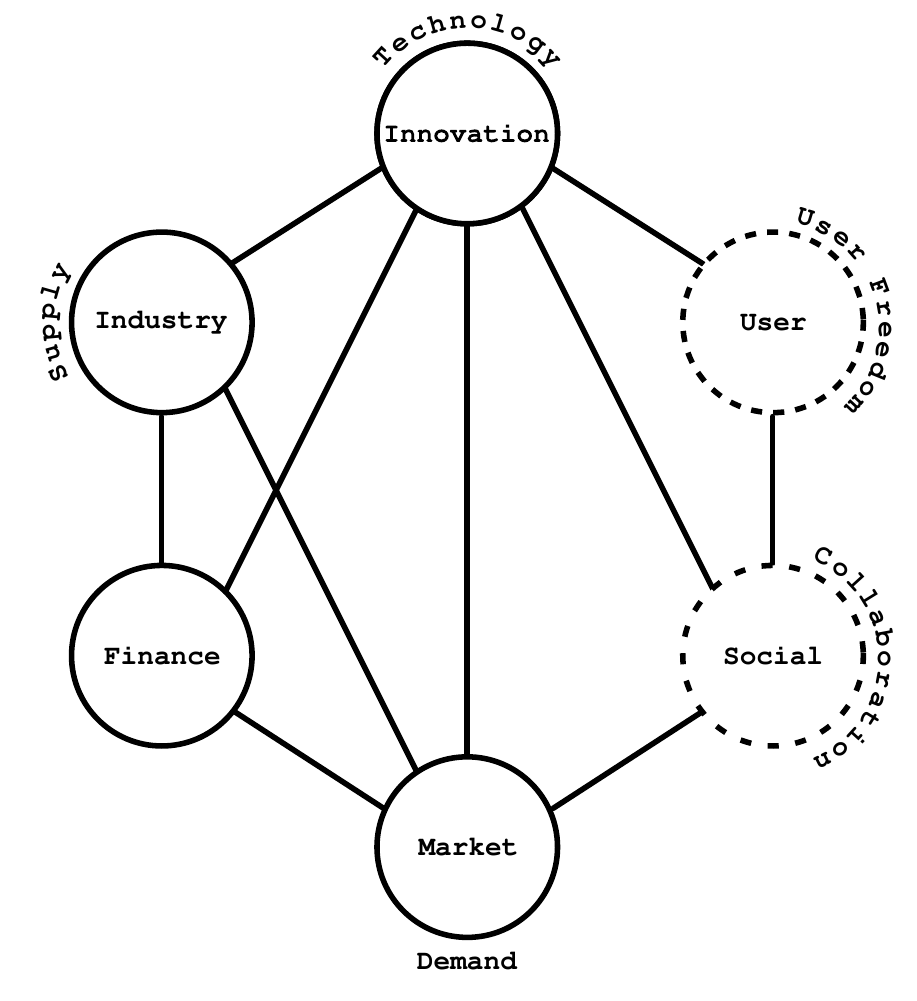}
	\caption{Redepiction of perspectives from the balanced scorecard as reported in \cite{camponovo2003business} along with the hidden perspectives}
\end{figure}

This demands the users to be viewed as intellectual resources, who could develop, produce, use products and not only as conventional consumers of products. On the whole the two perspectives have not only created a community approach in place of the traditional internet business model, but also aims at maintaining the structure and properties of internet in using it as a free medium to share, collaborate transparently. This points out the problem associated with traditional analysis of internetworked business models.


\subsection{Displacement and Bifurcation}

With rapid increase in service centric networking, the communication between the users, communities were dominated by centralized services which demands the user to be subscribed to enjoy communication, conversation, and social networking services. Inclusion of free of cost service attracted a lot of users to subscribe for the required service by submitting their personal details and may even extend to log their own conversations, involving personal events, information etc., in a remote system\cite{alternet2014sarah}.

Advertisement model have powered the categorization of users which adversely affected the structure of the internet. With the advances in computing and networking technology, fuzziness in boundaries between the users, communities and service providing organizations in internet are being eradicated. This resulted in creation of islands for each player with distinct practice in sharing and communication while amplifying the interlocked dependency between them. The increase in distance between the user of the service and the provider of the service also amplified the distance between data and analysis, while tightening the dependency between them and obscured the policy from mechanism.

With the evolution of such structure the problems associated with it also got more concnetrated between the boundaries of players of the internet. Sprouting at the interface of the players are the problems of dependency lockin and centralization respectively, which significantly influence the structure of internet.

\begin{figure}[h]
	\centering
	\includegraphics[width=0.90\linewidth]{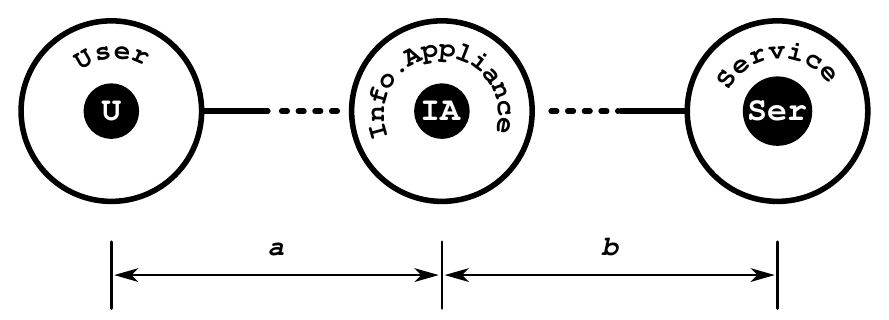}
	\caption{IA induced bifurcation between User and the Service}
\end{figure}

A simpler method mechanised by centralized web services was shown in Fig.2. It shows that the IA\textit{(Information Appliance)} which is placed in between the user and service increases the accessibility to the HMI\textit{(Human Machine Interface)} while tightening the users computing dependency $\mu$ -- over the service through the IA. The distance between the user and the information appliance $a$ -- keeps on decreasing because of the portable advantage and ever increasing technological miniaturization in manufacturing IA's. This results in several kinds of ubiquitiously networked devices configured for useful applications such as home automation and management systems, personal digital assistants, wearable health monitoring devices, etc.

On the other hand, the logical distance $b$ -- between the information appliance and the service is widening in centralized web based business model implementations to access the necessary computing resource required to operate on the information collected through the IA. This model reduces the information appliance to an integrated set of HMI and software applications bridging the data transfer between the user and the service via internet. 

On the whole, the displacement of the service requested from the user depends upon the number of web based centralized services and logical distance between them. With the availability of different kind of web services, computational dependency increases in number and leaves the user with just the IA. 
\begin{figure}[h]
	\centering
	\includegraphics[width=0.95\linewidth]{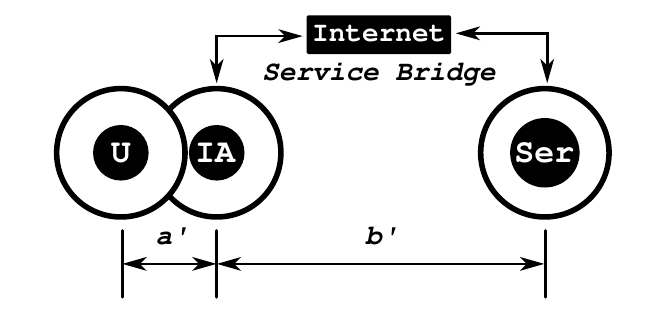}
	\caption{Bifurcation in Computing between the  User, Information Appliance \& subscribed Centralized web Service}
\end{figure}

Fig.3 illustrates such an example of an user subscribed to a service interfaced through the IA. It depicts how such practice can result in a bifurcated structure of computation and how major role played by internet in it as a \textbf{\textit{service bridge}} amplifies the displacement. The distance $a'$ -- shows that the actual distance between the user and the IA is smaller than the distance \textbf{[$b' = \delta + b$]} between user and the service, where $\delta $ is the bridge distance. Since the bridge distance represents the connectivity between the IA and the service through internet, which is not a real distance, actually influences the user to use the service wherever internet service is available, but mostly acting as a bridge for centralized web services instead of supporting collaborative or peer to peer communication. This reduced form of internet as a service bridge amplifies the bifurcation phenomenon such that the real computation over data happens in a remote system and the computation power in the information appliance left only for producing and using richer HMI and service locked applications. The increasing concentration of the centralized web services with time replacing all the necessary localized computing applications, may naturally modulate the distributed nature of internet into a service only bridge between the services and the user.
\begin{figure}[h]
	\centering
	\includegraphics[width=\linewidth]{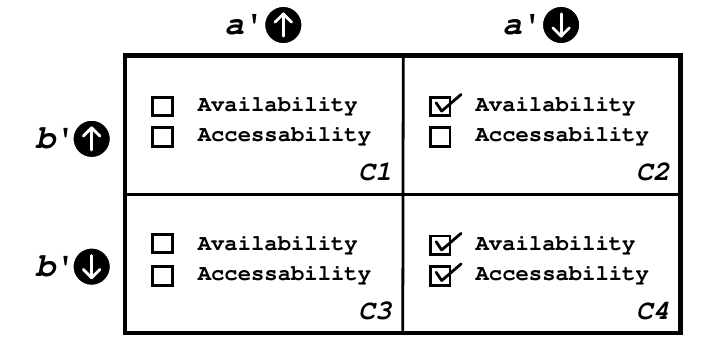}
	\caption{Illustration of relation between distance and dependency}
\end{figure}

Fig.4 illustrates how the distance between the user, IA and service influences the availability and accessibility(\textit{localized accessability}) of computation resources and the dependency between them resulting in different web based business models. As shown, four possible models \textbf{C1}, \textbf{C2}, \textbf{C3}, and \textbf{C4} can be generated based on the combinations of the distance $a'$ and \textbf{[$b' = \delta + b$]}. The cases C1 and C3 does not offer any attractive computational leverage because of the unavailability of the computing device itself(increased distance $a'$). On the other hand the cases C2 and C4 elucidates two polarized business models - centralized and distributed respectively. With decreased $a'$ and increased $b'$ number of web based centralized services increases in concentration. With the decreased $a'$ and $b'$, concentration of collaborative and peer to peer sharing model would increase. Thus the $b'$ distance influences the dependency \textbf{$\mu$} in a directly proportional way.

\begin{equation}
\mu \propto \textbf{$b'$}
\end{equation}

Equation 2 shows how dependency gets amplified (from $\mu$ to $\mu'$) in a centralized web service model with the number of subscribed users represented by the parameter $n'$. 
\begin{equation}
\mu' =  \textit{ f } (a', b', n')
\end{equation}
In an ideal distributed model the distance parameters  reduces to zero, essentially eradicating the dependency on centralized web services. Such a situation will redisplace the computing part back to the user and equip localized IA with HMI, necessary applications and tools, facilitating peer to peer and collaborative information management throughout the internet.

\subsection{Perpetual Lockin Environment} 

With the networked advantage of the internet, the requirement for recommendations and suggestions from the service have grown among the users which demands considerable storage and computational power. The organizations which does information brokerage took advantage of this scenario and offered the analytics as service for the user by consuming data from the user\cite{handleyp2p,pasquinelli2009google,jakobsson2010pirates}. 

\begin{figure}[h]
	\centering
	\includegraphics[width=0.75\linewidth]{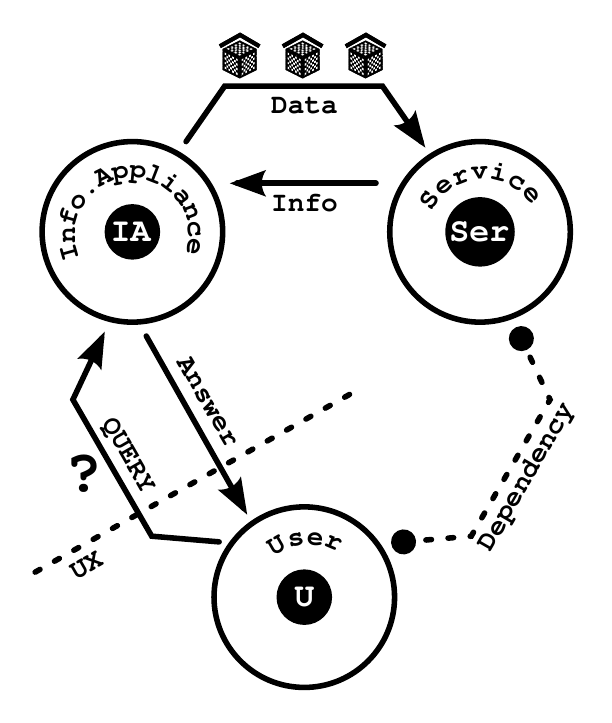}
	\caption{User - Appliance - Service Lockin}
\end{figure}

To sustain the reduced competition, organizations providing the analytical services for the users also design, manufacture, market and sell information appliances and software applications which collects the data from the user end. Information appliance converts the users requirments into queries understandable by the centralized service. These queries along with the data are then sent to a centralized analysis engine which acquires the data from all subscribed users channeled to their centralized analytics and datastore system, through internet and performs the necessary analysis over the logged data and returns the required results or suggestions back to the users. 

Such a policy is depicted in Fig.5, has created a market of appliances, and market of services. Since the data has to be transferred between the appliance and service, the complete mechanism requires a high bandwidth internet connection. This creates a cost overhead for the service users and mandates them to always stay online, even if the internet bandwidth cost is reduced to fraction of its current value. This mandation has driven the user to subscribe multitude of other services offered by the service providers.

This scenario creates an invisible mutual dependency lockin situation between the service providing organization and the users as shown in Fig.6, demanding more and more storage capacity in the service side, resulting in more computational power requirement. Organizations which have selected the above model would be investing more for storage and computational power. This situation also presses the organization to maintain such facilities as the number of users subscribe to the service increases.

\begin{figure}[h]
	\centering
	\includegraphics[width=\linewidth]{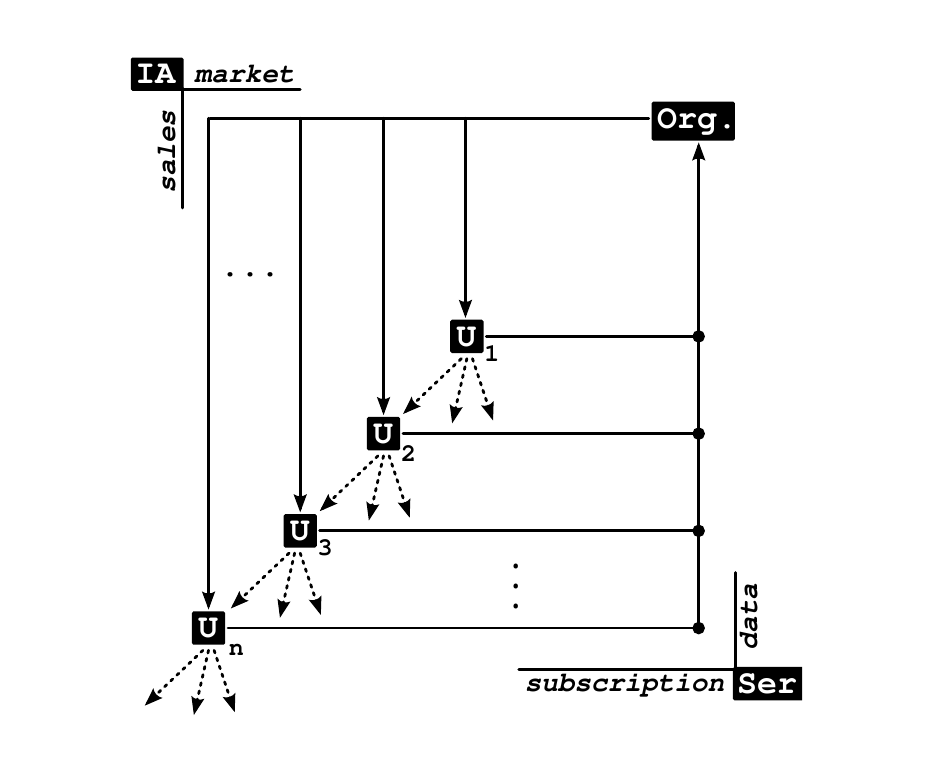}
	\caption{Perpetual Dependency Situation}
\end{figure}

Fig.6 plainly represents about how the organizations offer centralized web services targeted to specific community and user base, creating a \textbf{\textit{recursive attraction}} phenomenon that pulls more and more users perpetually to subscribe for the same service or any other similar competitive service. This phenomenon actually propels the marketing coverage not only by targeted advertising but also makes users to recommend the service to other users. Such recursive attraction based business model leads to \textbf{\textit{perpetual dependency}} which becomes hard to break pressing a new paradigm shift in mechanism and policy of the implemented technology with which the same service can be accessed without sacrificing freedom in knowledge building and sharing.

As explained in II.B, the above phenomenon accelerates users to use internet as service bridge alone rather than a complete medium, 
which resulted in analytics based baits for consumer lock-in that leads to fracturing of the network into large opaque consumer blocks farmed by these organisations. While this business model is justified in the current platform with its inherent division between data providers and data consumers, it can be made redundant and hence unnecessary. This has resulted in a mutually inseperable and irreversible lockin between the service seeker and service provider.

\begin{figure}[h]
	\centering
	\includegraphics[width=1.06\linewidth]{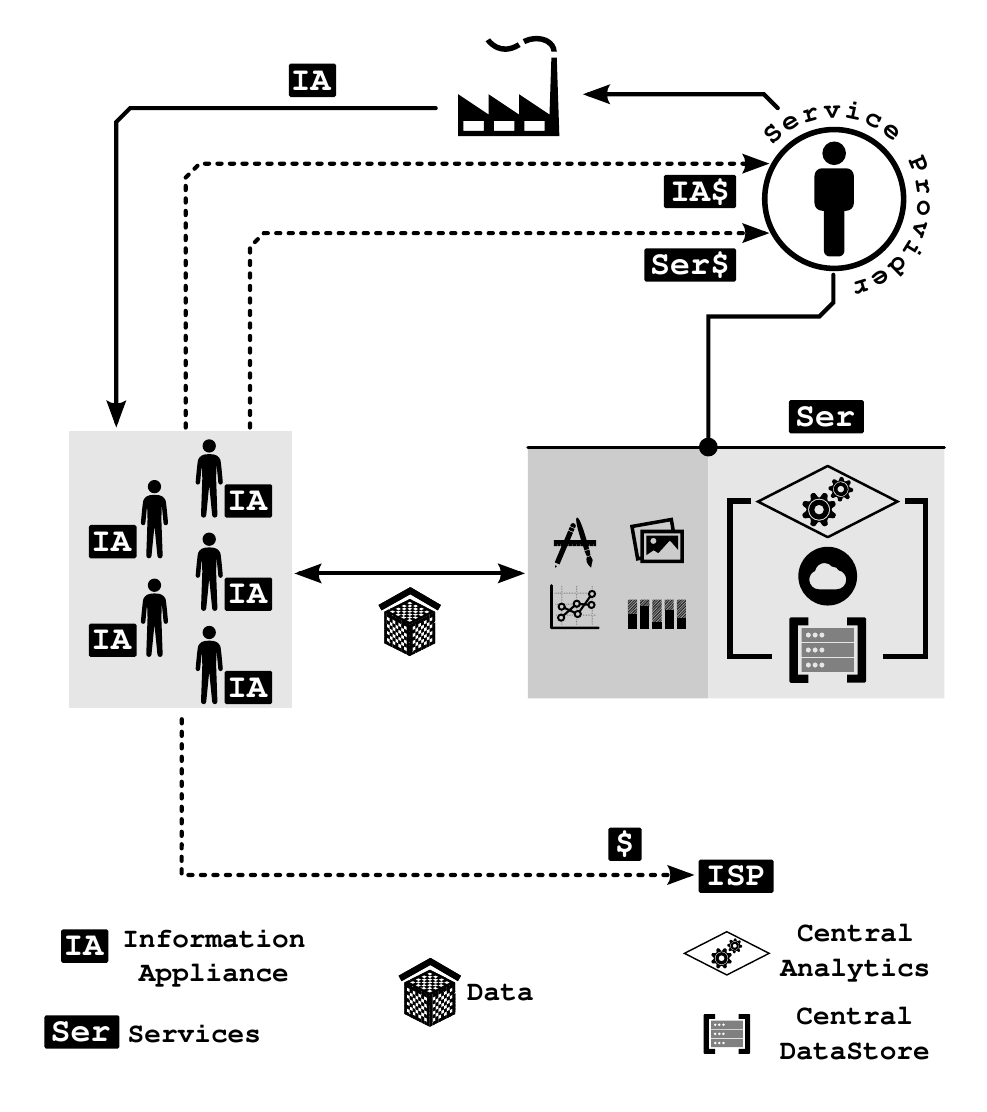}
	\caption{Lockin mechanism in Centralized Web Service based Business Model}
\end{figure}

The centralized business model which injects user-appliance-service lockin policy actually subserviates both the service provider and the user to continuously depend deeply on each other. A detailed qualitative model of such a business practice is shown in Fig.7. The figure depicts only an approximate mechanism how the model works. Based upon the market research, the service provider creates a remote online service and markets it throughout internet targeting the communities and users closely matching the requirement. For the subscribed user the organization provides both the appliance and service, around which the whole mechanism operates. 

With the increase in number of centralized services, tools and applications that can be packaged and distributed to the end user are now being marketed as centralized service, mandating the user to stay connected with the remote service through internet. Moreover when the centralized tool is proprietary and does scientific, engineering, or statistical computation upon the data provided by the user\cite{tellspecweb,scioweb,healbeweb}, then the mathematical knowledge behind the mechanism of the service becomes remote. This situation is much worse than a proprietry tool used locally for a specific purpose, because not only the knowledge become obscured but also placed remotely.

Even if the service is affordable, it adversely affects two other vital factors of knowledge propagation - availability and accessability, added with all the adverse impact of equivalent proprietary tools and patented techniques in the computing domain. Because of this process the user or a community, locked in with the service may get transformed into data acquisition instrument for the service and becomes ignorant about the mechanism and policy fabrics of the service.

\subsection{Vulnerable Chain Links}
Web article on Free Speech over internet\cite{articleefffswl} by Electronic Frontier Foundation(EFF), elucidates how the dynamic structure of internet are vulnerable at the intermediary points in the linking chain such as Web Hosting Services(WHS), Upstream Providers(UP), Domain Name System(DNS), Internet Service Providers(ISP), Search Engines(SE), Payment Service Providers(PSP), Third-Party platforms(TPP) as shown in Fig.8.

\begin{figure}[h]
	\centering
	\includegraphics[width=\linewidth]{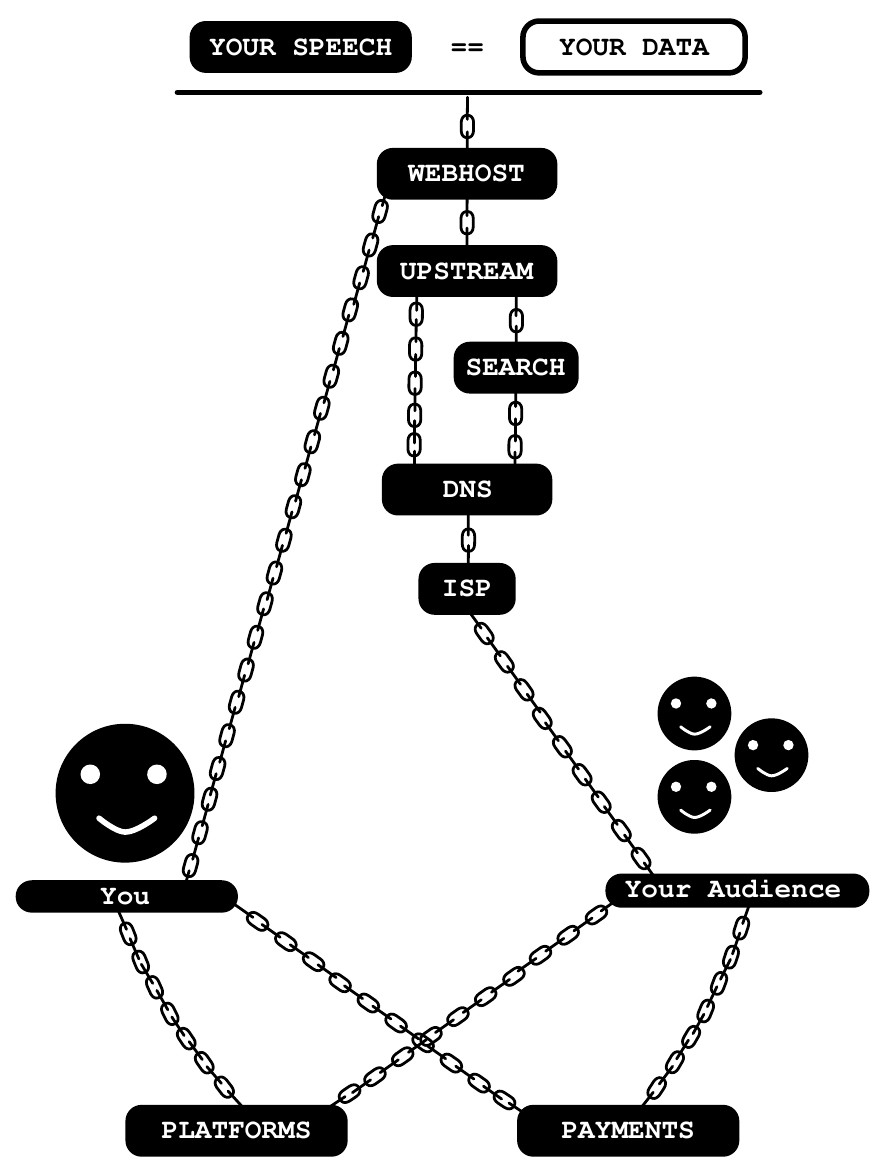}
	\caption{Vulnerable Intermediary Chain Links in Internet - redepicted from \cite{articleefffswl}}
\end{figure} 

Similar to free speech, the commons generated data that are vital for free data propagation, collaborative ranking, information sharing, creation of tools in both hardware and software, research and development also suffers from the same vulnerable intermediaries. Increase in the centralized web services proportionally raises the scope of vulnerabilities between the users who needs to share the resources and those who needs to access it. 

Furthermore, when free speech is substituted or appended with collaborative data operating on a free of cost service every user becomes both an author and audience. When the distance between the centralized service\textit{(processing and storing data)} and user\textit{(author and audience)} is increased, the possibility for censorship expands by pressing take down request from top hierarchy upstream providers which may adversely impact the users from accessing the data and the service. The negative impact further amplifies when the number of users were dependent on free of cost TPP\textit{(providing platform as service)} offered by the service provider. Thus when a self regulated knowledge building ecosystem is to be pragmatically implemented, the implementation model should be aware about those vulnerabilities listed out by EFF. Such awareness will help the developers to circumvent the vulnerabilities of the present structure of internet or atleast plan for an alternative method for the same. 

The conceptual model explained in this paper is assumed to work in a optimally distributed internet without any centralized intermediary between the commons. Thus it necessiates that every user in the model participating for collaborative knowledge building ecosystem to be equipped with essential tools and services that make their computing node distributed without compromising their privacy and anonimity. Even if such tools and services are not scalable for present day personal computers or information appliances, collaborative development will produce scalable tools in the near future.

\section{Mutual Sharing and Collaboration Model}

With the identification of problems associated with the web based centralized business models and the policies related to monetizing it, community based movements have developed alternative solutions that can adapt and evolve with the changes in the growth of internet. Solutions were built to eradicate mutual lockin dependency between the players of internet while sharing a symbiotic interrelationship between the end nodes. Several communities organize and regulate themselves using a model similar in structure as depicted in Fig.9.

Many collaborative organizations and communities such as Free software foundation\cite{fsf}, Wikipedia\cite{wiki}, Publiclab\cite{publiclab}, Citizen Science Projects\cite{citizenscipro}, Hackaday\cite{hackadayio}, Barefoot college\cite{barefootcol} have adopted and customized the model according to their needs. The goal of such a model is to always achieve self governing knowledge building ecosystem with transparency and freedom. Since the necessity of each community is different, diversed mechanisms are innovated to subsist the system financially and socially. Customization depends upon the philosophy, vision and tools used to build a pragmatic system from the model. 

The hardware and software tools used to build the system to solve a problem uses \textbf{\textit{democratic licenses and labels\cite{copyleft,gpl,ccl,ohl,cernoshl,ohanda}}} that supports knowledge propagation without any restriction, and encourages to distribute and share the associated resources publicly. Tools which carry democratic licenses and labels, self propagates the tools to be used, studied, edited and redistribute in the same way how they have attained it. So any changes or customization made when published will result in an avalanche adaption, peer review and deployment. Such license structure offers alternative advantages as comapared to the proprietary licensing system, as listed below :
\begin{itemize}
	\item Anybody can participate in technololgy development.
	\item Commons who do not write code or design hardware, can contribute for documentation on how to use the tools and products.
	\item Technology development becomes a social activity.
	\item VCS\textit{(Version Control System)} aids in collaborative creation and management of high quality tools.
	\item Designs can be customized to suit the need.
	\item Malicious code inclusion can be eradicated just by peer reviewing.
	\item Ensures four\cite{fourfreedoms} essential freedoms to be propagated as recommended by democratic licenses.
	\item Encourages localized diversified businessess, without any subversive lockin situation eradicating natural monopoly.
\end{itemize}

The sharing model takes advantage of the personal computing power available with the end nodes of internet(\textit{end users}), networking techniques, encrypted data sharing and communication mechanisms. It aims in replacing the central analytics and datastore services with potential distributed analytics and datastore tools which can be made available in every personal computer. In order to improve the quality of such a system respective hardware and software tools, instruments, data and documentation are developed and shared collaboratively using peer to peer sharing tools and peer reviewing mechanisms. Again the peer reviewing mechanisms may differ between communities.
\begin{figure}[h]
	\centering
	\includegraphics[width=1.02\linewidth]{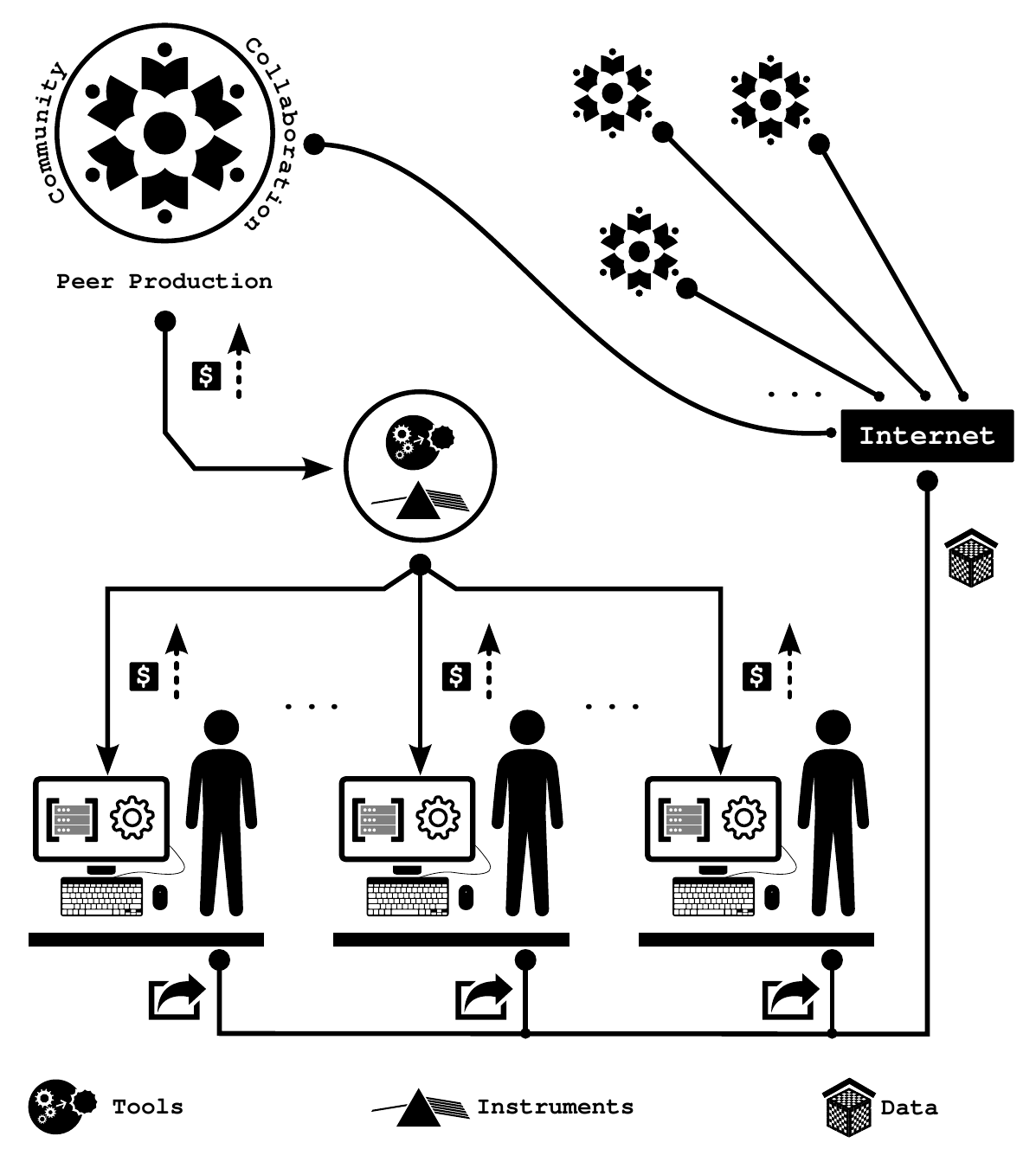}
	\caption{Solution through Distribution and Collaboration}
\end{figure}

The flow diagram in Fig.9. elucidates a cyclic flow of knowledge between the collaborative community and the end user. The community develops the necessary hardware, instruments, software tools, guidelines, documentation required to use the product in the end user system. The hardware and software development, debug, release cycle are regulated through collaborative distributed VCS\cite{distverconsys,gyerik2013bazaar,de2009software}. The user sets up the product as per the documentation, and starts generating data in a free format from their personal computer. The data, analysed information, and reports are then shared back to the community(\textit{without compromising privacy}) to better understand the problem, identify the pattern, learn, discuss, evaluate and rank a practical solution. Sharing involves scalable volume of data, which are shared using peer to peer file sharing tools like retroshare\cite{p2ptoolretroshare,rogers2010private}. While identifying a solution based on the data acuired may not be required for every community, it is essential for learning, hacking, research and development communities which is vital for knowledge propagation and innovation. 

Systems that require good data acquisition, analysis, report generation and decision making are needed for quality evaluation of physical, chemical, biological phenomena. Even if scientific and engineering methods are available to solve such problems, most of them are either proprietary or available as remote service with patented analysis algorithms. Furthermore, having such a system in every personal computing environment empowers the user to monitor and assess their personal health; evaluate the quality of their local environment such as their homes, gardens, agricultural lands to name a few, with complete transparency and freedom.
With such a system counterfeiting of food and related products can be rooted out. It also acts as a shield for the user to gaurd against the false advertisements with which the present advertisement model suffers from selling bad quality, adversely affecting adulterant or counterfeited products. Every user can deploy a collaboratively ranked and decentralized quality auditing inspection system, distributing the inspection ability to public that pragmatizes the ideal anticounterfeiting system and false advertisement watchdog system.

To attain such profound impact, freedom and transparency in development and usage of measurement tools, mechanisms, policies cannot be compromised. In order to realize the model practically, every node must be equipped with tools built with following important properties :
\begin{itemize}
	\item No bifurcation between the information appliance and the computational services.
	\item Collaborative development and sharing of required hardware design, software tools and data using democratic licenses.
	\item Collaboratively ranked distributed analytics engine.
	\item Distributed datastore and data management.
\end{itemize}

Participation of commons in the collaborative peer production and ranking using free software and hardware tools is vital in deciding the effective impact created by the products inherited with these properties.

\subsection{Commons Based Peer Production}
As defined in Wikipedia, CBPP\textit{(Commons Based Peer Production)} describes a new model of socio-economic production in which the creative energy of large numbers of people is coordinated into largescale, meaningful projects mostly without traditional hierarchical organization. The model focuses on bridging commons to use the data generated from the built tools, devices and instruments and share it with others which aids in self regulation and governance by building knowledge collaboratively from the information obtained. It is also necessary to share how the tools, devices and instruments were built that helps the commons to build whatever necessary by reusing them and contribute back to the community recursively.

Commons based projects like the GNU, Wikipedia, GNUNet, Freenet, GNU Social, MediaGoblin, Movim, Yacy, ArkOS, NounProject, OpenClipArt have proven there usefulness and impact in enabling federated  communication and sharing between individuals. Their success along with the heightening of privacy concerns have provided a strong impetus for accelerating the decentralisation of the internet and its network infrastructure. Work on creating distributed/decentralised social networks\cite{alternet2014sarah,decentsocnet,yeung2009decentralization,breslin2007future,openpds,indiephone}, currencies, media publishing, chat applications, forums and development of distributed network infrastructure using wireless mesh networks\cite{commotionwireless,openwireless} is rapidly being carried out, tested and deployed by the community movements. With the open data movement helping simplify the tools\cite{anokwa2009open,opendatakit} necessary for opening up data\cite{opendatacommons,openfoodfact} and datastores wherever possible. 

In particular projects like MobileECG\cite{mobileecg}, OpenBCI\cite{openbci}, Spectruino\cite{spectruino}, Publiclab - DIY Spectrometer\cite{publiclabspectrometer}, Hackaday - RamanPi\cite{ramanpi}, LaserSaur\cite{lasersaur}, GNURadio\cite{gnuradio,tucker2009prototyping}, FSAE Electric Race car\cite{fsae2014electricracecar}, OpenSource Lab\cite{appropopensourcelab,pearce2013open}, Sensorica\cite{appropsensorica,sensorica}, SatNOGS \cite{poblet2014telecommunications}, AltOS \cite{telemetrum,altusmetrumaltos}, OpenPilot\cite{openpilotwiki,bricenovirtual}, Paparazzi\cite{paparazziurl,hattenberger2014using} etc., developed with available free software and open hardware tools, have great impact in education, research and development, were licensed democratically and proved to improve the quality of development by collaborative contribution. 

It is clearly observable that these projects do adapt the principles specified by CBPP in their development cycle. Even when some projects were not financially successful in their business the avaialability, accessability and collaborative nature of internet have helped the communities around the world to reuse such existing projects and customize them according to their needs, while attributing the founders of the project and supporting them back financially which may regenerate the projects to become financially stable and successful. This nature would be nearly impossible in traditional centralized business model. If the startup fails in financial or marketing terms, the kowledge behind the project too would not be propagated and get terminated along with the business.

CBPP will not only accelerate the propagation of real knowledge behind a startup, but also powers the commons to verify whether the promised goal can be achieved, help them analyze the mathematical mechanisms and even suggest to improve them through collaborative contribution. Furthermore, this feedback mechanism will act as a watchdog that self regulates the quality of startups in crowd sourced business environments. Crowd sourced projects like WeIO\cite{weio}, RedPitaya\cite{redpitaya,openelectronicsredpitaya} which already follow such mechanism have proved to be both financially and socially successful. Startups which follow the centralized and closed model suffer from their complete dependency on marketing strategies to sell their product. This presses such startups to concentrate and invest more in marketing and sales strategies than in other areas of their business model.

With the longitudinal research, results and open data collected and analyzed peer production surveys\cite{surveyofpeerproduction}, which itself is collaboratively maintained depicts that CBPP will be the next state of production, governed by the commons and not by the centralized markets. Collaborative wikis like elinux\cite{elinuxwiki}, appropedia\cite{appropediawiki} driven by common interests have brought together diversed tools and technologies, people with different skills which resulted in creation of low cost alternative solutions for centralized, proprietary products. This enables commons to think globally and act locally in distributed manufacturing and production by adapting the principles of CBPP.

Thus in addition to conceptualize and build a non-hierarchial, distributed knowledge building ecosystem model, it must also comply with the potential goals\cite{cbppwikiprinciples} defined by the commons based peer production as mentioned below :
\begin{itemize}
	\item \textbf{\textit{Modularity}} - Facilitates collaborative development by dispersing the objectives into seperate modules.
	\item \textbf{\textit{Granularity}} - Scales the commons contribution in development of individual modules.
	\item \textbf{\textit{Low cost Integration}} - Producing a complete quality controlled product with collaboratively ranking of modules and integration mechanism at relatively low cost.
\end{itemize}

\subsection{Business Model Generation}
Creating a business model adapting the above elucidated principles is vital for recognizing the importance of propagation of knowledge. Measuring success of the generated business model in terms of profit will not alone suffice to evaluate its economic and societal impact. Thus the quality of generated business models should be measured with how well it is transparent and the support that it provides to enable knowledge propagation to the end users of the product. Based on CBPP several models were constructed and functioning all over the world successfully. Massimo Menichinelli research reports\cite{menichinelli2008openp2pdesign,menichinelli3,niessen2010open} states that collaborative business models generated in the CBPP domain such as fablabs and hackerspaces sustain both in economic and social form. As per the case studies listed in research reports of P.Troxler about how fablabs struggle and innovate business models\cite{troxler2010commons} to sustain themselves financially it becomes clear that the communities and commons try to perpetuate a innovative ecology in digital fabrication realm. Being replicated from the free software model, CBPP especially in fablabs and hackerspaces proved to be next revolution to reach democratic production of physical goods\cite{tanenbaum2013democratizing}. Similarly business models around opendata, distributed network architectures are necessary to connect, accelerate and sustain the CBPP based business, while eradicating the centralized control over computation and information.

\section{Conclusion}
Internet is no more a network of devices sharing information and other resources accelerating knowledge generation and business incentives. It has evolved into a platform where pragmatic impact on the society happens with public participation. Internet and World Wide Web digitally manifests the social system with unparalleled flexibility, freedom and transparency. It invariably constitutes systems that works with subjugated policy, made to work for yielding business profits as well as systems that work with liberty. We presented a significant review of how hierarchical internet based business models subjugated knowledge propagation and how available collaborative commons based peer production model that serves not only as an alternative to the hierarchical models. 

Creative individuals unite based on their common interest and evolve into communities innovating revolutionary ideas, implement them and generate new business models without compromising the liberty in propagating the knowledge associated with the project. Such a model allows anybody to participate in hacking, researching and producing ideas and projects from where the potential support from the public leverage together to solve their individual problems through which a common societal problem can be rapidly addressed and solved.

We believe that with the advent of distributed tools and systems, CBPP model would further propagate into every computing appliance replacing the centralized and hierarchical business models, preserving the true and neutral nature of internet by criticizing and adapting along with the changes occuring in the internet's own timeline. 


\bibliographystyle{unsrt}

\end{document}